\documentclass[12pt]{article}
\usepackage{epsfig}
\usepackage{graphicx}
\begin{document}
\begin{center}
{\large \bf A unique parameterization of the QCD equation\\[1mm]
of state below and above $T_c$}
\vskip 5mm
M. Bluhm$\,{}^{a)}$, B. K\"ampfer$\,{}^{b)}$\footnote{
Speaker at XLII Winter Meeting on Nuclear Physics, Bormio, It., Jan. 25 - 31, 2004}, 
G. Soff$\,^{a)}$
\vskip 5mm
${}^{a)}$ Institut f\"ur Theoretische Physik, TU Dresden, 01062 Dresden, Germany\\
${}^{b)}$ Forschungszentrum Rossendorf, PF 510119, 01314 Dresden, Germany
\end{center}
\vskip 5mm

\begin{center}
\begin{minipage}{130mm}
{\small
We present a unique parameterization of the equation of state of
strongly interacting matter in the temperature interval 
$0.6 T_c \cdots 3 T_c$ at $\mu = 0$ 
within a quasi-particle model based on quark and gluon degrees of freedom.
The extension to non-vanishing baryon-chemical potential is discussed.}
\end{minipage}
\end{center}

\section{Introduction} 

The equation of state (EoS) of strongly interacting matter contains
important information about bulk properties and the thermodynamics. In
fact, one needs the EoS as input for describing hydrodynamically the
evolution of the early universe, dynamics and static characteristics
of neutron stars as well as certain stages of relativistic heavy-ion
collisions. Considerable progress has been achieved during the last
years in calculating the EoS from first principles resting on quantum
chromodynamics (QCD). By advanced sampling techniques the EoS of
strongly interacting matter is numerically accessible from lattice QCD
calculations. Many details have been considered for
particle-antiparticle symmetric matter. Here, at a certain value of
the temperature, $T_c$, the susceptibilities corresponding to the
Polyakov loop and the chiral condensate display pronounced peaks
indicating deconfinement and chiral symmetry restoration (cf.\ \cite{Kar1}). 
Above $T_c$, in the deconfinement region, the relevant degrees of
freedom of strongly interacting matter are thought to be quarks and
gluons. Despite asymptotic freedom, in the range of $T_c < T < 2.5 T_c$, the
EoS behaves highly non-trivial, preventing a direct perturbative
treatment. There are various attempts to understand the EoS in this
range and above, such as dimensional reduction, resummed HTL scheme,
$\Phi$ functional approach, Polyakov loop model etc.\ (cf.\ \cite{Pei1}
for a recent survey). A controlled chain of approximations from full
QCD to some analytical expressions describing the lattice data without
adjustable parameters would be of desire. Success has been achieved
for $T > 2.5 T_c$ \cite{Bla1}. In contrast, the range $T > T_c$, in particular
near to $T_c$, is covered by
phenomenological models \cite{Pes1,Kam1,Pes3,Lev1} with parameters adjusted to
lattice QCD data.

Below $T_c$, in the confinement region, hadrons are thought to be the
relevant degrees of freedom. Detailed information about the EoS in this range
from lattice QCD became available very recently \cite{Kar2}, offering a
chance to check phenomenological models for the first
time. In \cite{Kar3} lattice QCD data from \cite{Kar2} for $T < T_c$
have been shown to agree with results of a hadron resonance gas model
including a large number of states up to 2 GeV. Taking a
different point of view, we present here a description of the lattice QCD
data \cite{Kar2} in the range $0.6 T_c < T < 3 T_c$ by our quasi-particle
model of quarks and gluons \cite{Pes1,Kam1}. Up to now it has been proven 
that our model can reproduce lattice data in the range 
$T_c \cdots 3T_c$ for the pure gluon plasma \cite{Pes1} and for quark-gluon
plasmas with various quark-flavor numbers \cite{Kam1}. Here, our focus is
to cover the lattice QCD data in the confinement region.

Our paper is organized as follows. In section 2 we recapitulate features
of our quasi-particle model. The comparison with lattice QCD data at vanishing
chemical potential is performed in section 3. We discuss the obtained
parameterization of the equation of state in section 4. Section 5 is devoted
to non-vanishing chemical potential. The summary can be found in section 6.

\section{Quasi-particle Model} 

Our quasi-particle model of light and strange quarks ($q, s$) and gluons ($g$) 
is based on the entropy density
\begin{equation}
\label{e:ent}
s = \sum_{i = q, s, g} s_i, 
\,
s_i = \frac{d_i}{2\pi^2T}
\int dk k^2 \left[ \frac{\left( \frac{4}{3}k^2 + m_i^2 \right)}{\sqrt{m_i^2 + k^2}} 
(f_+(k) + f_-(k))
- \mu (f_+ (k) - f_-(k)) \right]
\end{equation}
and the quark number density
\begin{equation}
\label{e:den}
n = \frac{d_q}{2\pi^2} \int dk k^2 \left[f_+(k) - f_-(k)\right]
\end{equation}
with degeneracies $d_q = 12$, $d_s = 6$ and $d_g = 8$
as for free partons, and distribution functions  
$f_{\pm}(k) = (\exp( [\sqrt{m_i^2 + k^2} \mp \mu]/ T) +S)^{-1}$
with $S = + 1$ ($- 1$) for quarks (gluons). The 
chemical potential is $\mu$ for light 
quarks and anti-quarks, respectively, while $\mu = 0$ for strange quarks and gluons.
The parton effective masses
\begin{equation}
m_i^2(T,\mu) = m_{i,0}^2 + \Pi_i(k; T, \mu)
\end{equation}
have a rest mass contribution $m_{i,0}^2$ and, as essential part, the 
one-loop self-energies at hard momenta $\Pi_i (k; T,\mu)$. The crucial point is to
replace in $\Pi_i (k; T,\mu)$ the running coupling by an effective coupling, 
$G^2(T,\mu)$
\cite{Pes1}. (As shown in \cite{Pes9}, it is the introduced modification of
$G^2(T,\mu)$ which allows to describe the lattice QCD data, while the use of the
1-loop or 2-loop perturbative coupling together with a more complete description
of the plasmon term and Landau damping restricts the approach to $T \ge 2.5 T_c$.) 
In doing so, non-perturbative effects are thought to be
accommodated in this effective coupling. 
In massless $\Phi^4$ theory such a structure of the entropy density emerges by
resumming the super-daisy diagrams in tadpole topology \cite{Pes4},
and \cite{Pes5} argues that such an ansatz is also valid for
QCD. \cite{Bla1} point to more complex structures, but we
find (\ref{e:ent}, \ref{e:den}) flexible enough to accommodate the lattice data (see below). 

The pressure reads accordingly 
\begin{equation}
\label{e:pres}
  p = \sum_{i = q,s,g} p_i - B(T, \mu), \quad
  p_i = \frac{d_i}{6 \pi^2} \int dk \frac{k^4}{\sqrt{m_i^2 + k^2}}
  \left[ f_+ (k) + f_-(k) \right], 
\end{equation}
where $B(T,\mu)$ ensures thermodynamic self-consistency, 
$s =\partial p / \partial T$, $n = \partial p / \partial \mu$
together with the stationarity condition
$\delta p / \delta m_i^2 = 0$ \cite{Gor1}. 
Note that eqs.~(\ref{e:ent}, \ref{e:den}, \ref{e:pres}) themselves
are highly non-perturbative
expressions: Expanding (\ref{e:pres}) in powers of the coupling strength 
one recovers the first perturbative terms.
(Higher order terms would probably need higher orders in $\Pi$). 
For more details on our model see \cite{Pes1,Kam1,Pes3,Pes9,Pes5}. 

\section{Comparison with lattice data at ${\bf \mu = 0}$} 

(\ref{e:ent}) represents a mapping of lattice data for $s(T)$ on $G^2(T)$; to
determine $p$ one has to fix an integration constant, say $B(T_c)$. 
As in the lattice calculations \cite{Kar2} we have used $m_{i,0} = x_i T$
with $x_q = 0.4$ for light quark flavors, $x_s = 1$ for strange quarks
and $x_g = 0$. 
We have found as a convenient parameterization
\begin{equation}
\label{e:G}
G^2(T) = \left\{
\begin{array}{l}
G^2_{\rm 2-loop}, \quad T \ge T_c
\\[3mm]
a - b T / T_c, \quad T < T_c
\end{array}
\right.
\end{equation}
where $G^2_{\rm 2-loop}$ is the 2-loop coupling
\begin{equation}
\label{e:G2}
G^2_{\rm 2-loop} = \frac{16 \pi^2}{\beta_0 \log \xi^2}
\left[ 1 - \frac{2 \beta_1}{\beta_0^2} \frac{\log (\log \xi^2)}{\log \xi^2} \right]
\end{equation}
with $\beta_0 = (11 N_c - 2 N_f) / 3$,
$\beta_1 = (34 N_c^2 -13 N_f N_c + 3 N_f /N_c)/6$, 
and the argument $\xi = \lambda (T - T_s)/T_c$.
The parameters are $T_s = 0.80 T_c$, $\lambda = 7.6$,
$a = 359.0$ and $b = 334.5$ for $N_f = 2 + 1$, $N_c = 3$.
The comparison of our model with the lattice data \cite{Kar2,Kar3}
is exhibited in Fig.~1 by solid curves for entropy density and pressure,
respectively. One observes an impressively good description of the lattice data,
in particular also below $T_c$.

Remarkable is the change of the shape of $G^2(T)$ at $T_c$, see Fig.~2.
Above $T_c$, $G^2(T)$ resembles the usual 2-loop running coupling strength with a
regulator $T_s$, which ensures agreement with the data down to $T_c$.
Going down in $T$, at $T_c$ the growing of $G^2(T)$ is changed into
a moderate linear rise. 
Note that no order parameter is needed
which changes at $T_c$. This is possibly related to the fact that $s$ is
a measure for the density of states. Also, the degeneracies $d_i$ are
constant. 

Here we would like to contrast our quasi-particle model based on quark-gluon
degrees of freedom with the resonance gas model below $T_c$ \cite{Kar3}.
The authors of \cite{Kar3} state that many heavy resonances are needed to
describe the data (at $T_c$ the lightest hadron
contribution to the energy density is at the level of
$15\%$ \cite{Kar3}). Our result is in some correspondence, in the sense
that fairly heavy quasi-particle excitations are needed
which emerge from the large values of $G^2$.
One could also argue that a large number of excited hadron states at $T < T_c$
can be effectively described in an equivalent manner by a small number
of quasi-particle excitations, as we do here.  

\begin{figure}[ht]
\vskip 3mm
\centering
\includegraphics[width=5.1cm,angle=-90]{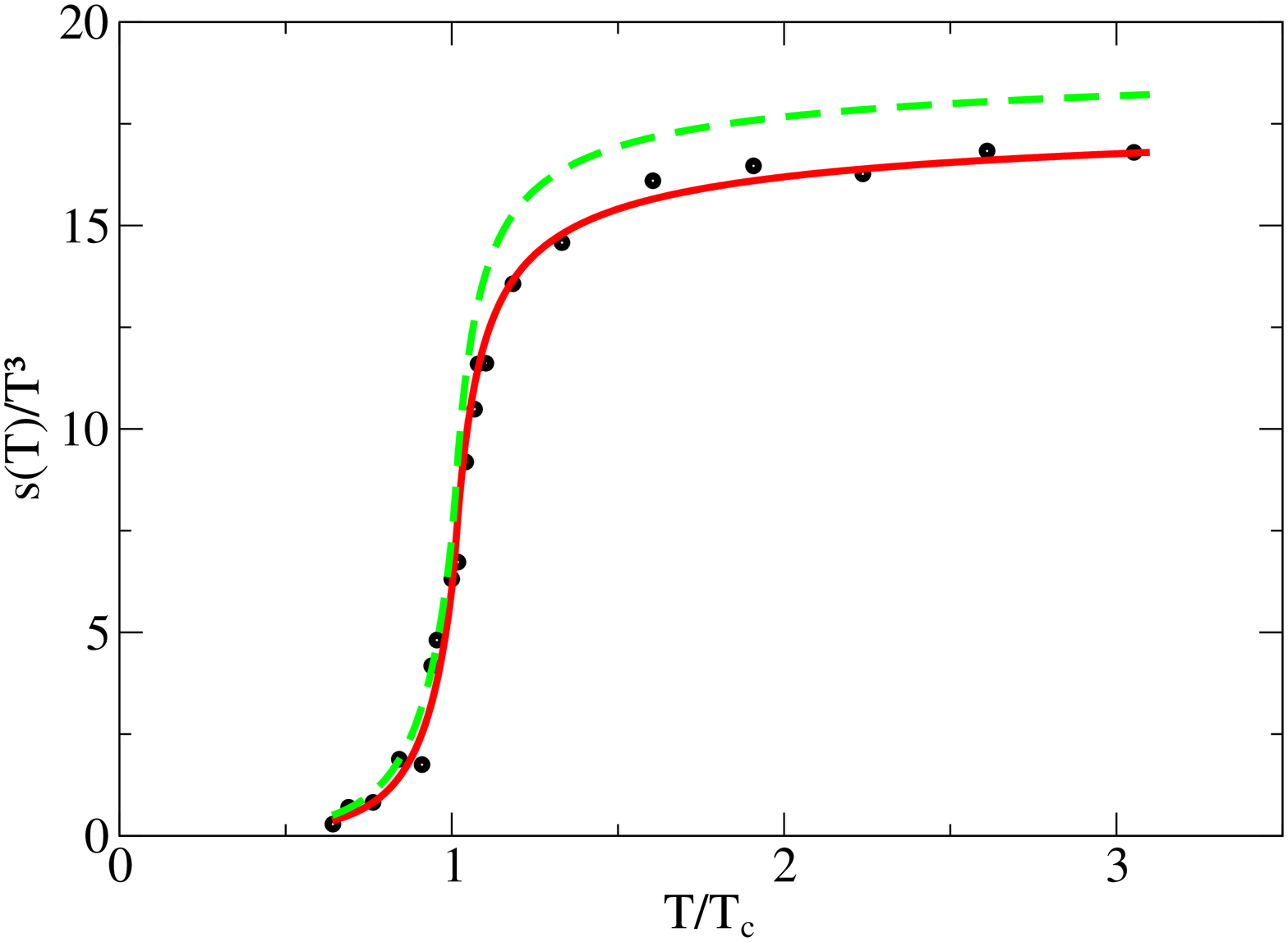}
\hfill
\includegraphics[width=5.1cm,angle=-90]{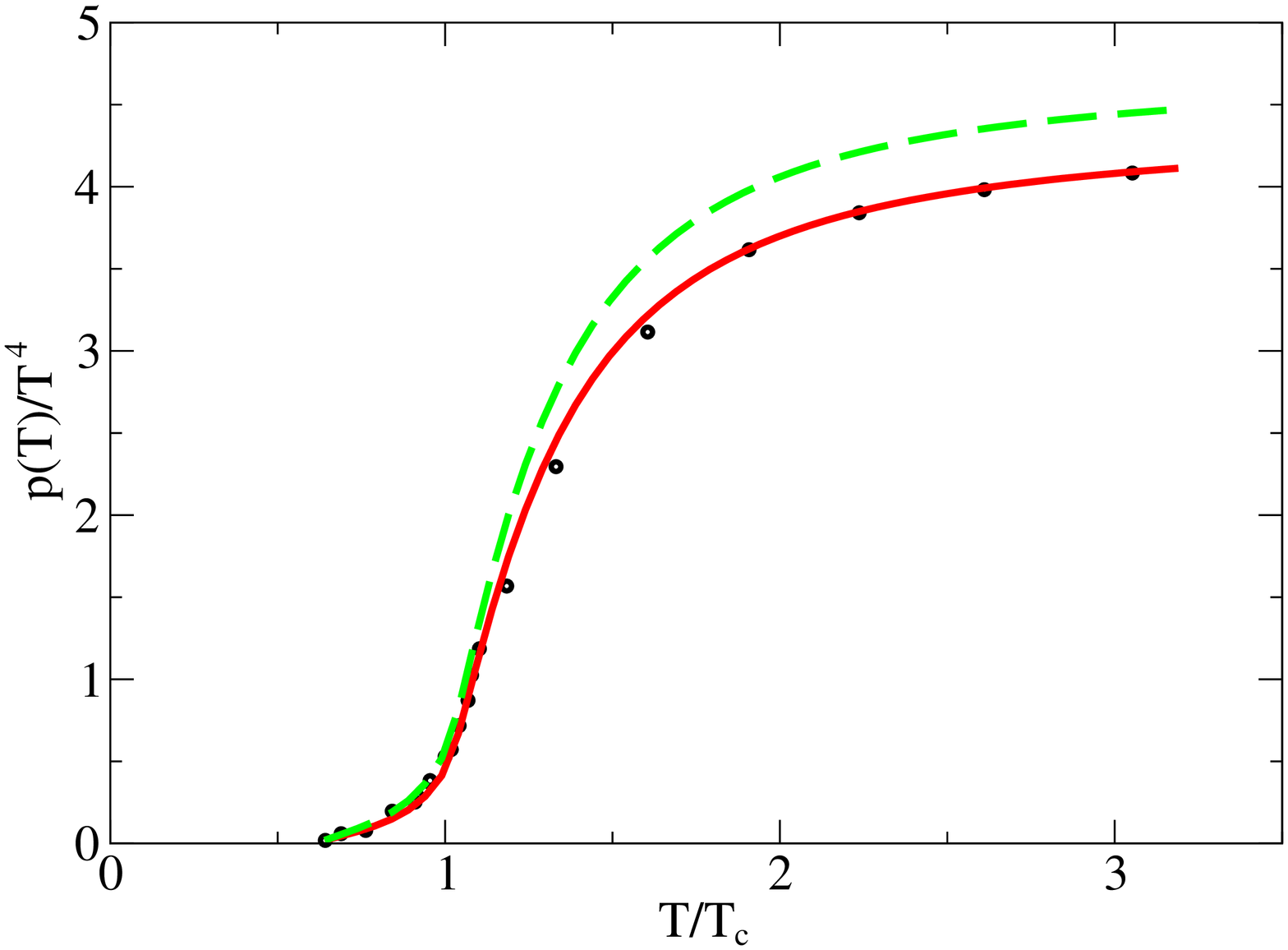}
\caption{
\label{fig:s}
{\it Entropy density $s$ scaled by $T^3$ (left panel) and pressure scaled by $T^4$ 
(right panel)
as a function of the temperature in units of the pseudo-critical temperature
$T_c$. The full (dashed) curves represent
calculations with lattice masses as described in text 
($m_{i,0}=0$ as an estimate of the chiral extrapolation).
Lattice data from \cite{Kar2,Kar3}.}}
\end{figure}

It is well conceivable
that, for determining the pressure (\ref{e:pres}), two distinct
integration constants would be needed, say $B(T_c^+)$ and
$B(T_c^-)$. However, one constant $B(T_c)=0.44T_c^4$ is sufficient to
describe the pressure, see right panel in Fig.~\ref{fig:s}. 

As mentioned above, the lattice calculations \cite{Kar2} use quite
heavy quarks. A possible estimate of the chiral extrapolation is to
put $m_{i,0}=0$ and repeat the evaluation of
eqs.~(\ref{e:ent} - \ref{e:pres}) with keeping the parameters in
(\ref{e:G}) fixed;  $B(T_c)$ is adapted to ensure a positive pressure.
The results are shown as dashed curves
in Fig.~\ref{fig:s}. Such a procedure is certainly too
simple. Only once lattice results for various
$m_{i,0}$ are at our disposal one can estimate the $m_{i,0}$
dependence of the parameters in (\ref{e:G}) and for $B(T_c)$ to achieve
a more profound chiral extrapolation.\\[1mm] 

\begin{figure}[ht]
\centering
\includegraphics[width=5.1cm,angle=-90]{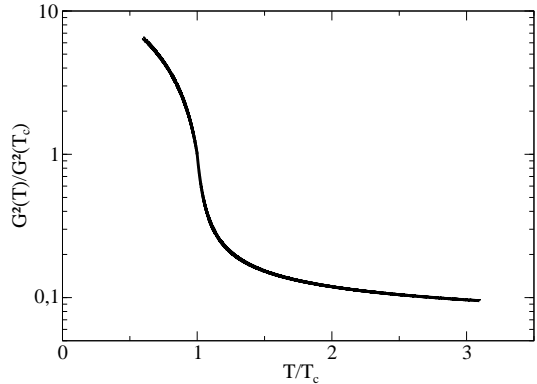}
\caption{
\label{fig:G}
{\it The effective coupling $G^2 (T)$ (in units of $G^2(T_c)$)
as a function of the temperature.}}
\end{figure}

\section{Discussion} 

Up to this point we consider
eqs.~(\ref{e:ent} - \ref{e:pres}) as a convenient unique parameterization
of the EoS, which one may make use of for purposes mentioned in the
introduction.
Further detailed lattice data are needed to test our model at even
smaller temperatures and to see where an explicit change to hadronic
degrees of freedom is required. 

In an restricted interval around $T_c$
one may speculate whether a quark-hadron duality is at work,
i.e. hadron observables are expressed in a quark-gluon basis and vice
versa. For example, \cite{Kar3} shows that in a narrow interval above
$T_c$ the hadron resonance gas model still agrees with the lattice
data. Above $T_c$, signals of hadronic states have been found as
well \cite{Tar}, an aspect advocated as important in \cite{Shuryak}.
Even more, \cite{Blaschke} has shown that a modified resonance gas model 
describes the lattice data above $T_c$ also successfully.

There are various examples of the quark-hadron
duality in the literature. We mention here QCD sum rules \cite{Shi}
which realize the duality, as discussed in \cite{Coh}. Another
observation is that in heavy-ion collisions the low-mass di-electron
spectra can be described by a quark-gluon plasma emission rate, even if the
space-time averaged temperature of the emitting system is below 
$T_c$ \cite{Gall}. For further aspects of the duality we refer,
e.g., to \cite{duality}.

\section{Non-vanishing chemical potential} 

With the advent of lattice QCD data at non-vanishing chemical potential
\cite{Fodor,Allton1,Allton2} the equation of state becomes accessible
in a large region. This is particularly important as the envisaged
CBM project at the future accelerator complex in Darmstadt \cite{CBM}
aims at exploring systematically the region of maximum baryon
densities at reach in heavy-ion collisions. 

As described in \cite{Pes3,Pes5} our quasi-particle model also covers
the equation of state for non-vanishing chemical potential.
The point here is that thermodynamical self-consistency allows to map
the lattice QCD data from $\mu = 0$ to $\mu > 0$ without further assumptions.
Indeed, as shown in \cite{Ungarn} our model agrees quite perfectly
with lattice QCD calculations at $\mu > 0$ for $2 + 1$ flavors. Here we focus
on the recent lattice QCD data for 2 flavors \cite{Allton2} with improved p4 action.
Given $G^2(T)$ (e.g., from the previous section, or from an analysis
of other lattice QCD data such as \cite{Allton2}) 
as boundary values, the partial differential equation
$a_T \partial G^2 (T,\mu)/\partial T + a_\mu \partial G^2 (T,\mu) / \partial \mu
+ a_{T,\mu} = 0$ with $a_T (\mu = 0) = 0$ and $a_\mu (T = 0) = 0$
(cf.\ \cite{Pes3,Pes5} for details) delivers as solution
$G^2(T,\mu)$, needed in $\Pi_i (k; T, \mu)$. Fig.~3 exhibits the characteristic
curves for solving this ''flow equation''. Asymptotically, $G^2$ is constant
along the characteristic curves. As discussed in \cite{Pes3,Pes5} we take
the characteristic curve emerging from $T_c$ at $\mu = 0$ (solid curve in Fig.~3) 
as indicator of the phase border line. It agrees up to intermediate values of $\mu$
fairly well with estimates based on lattice QCD calculations \cite{Allton1}.
The characteristic curves above the solid curve show a regular pattern
of sections of ellipses.
In contrast, the characteristic curves emerging from the $T$ axis below
$T_c$ are flatter and cross the former ones at larger values of $\mu$. 
This may indicate
that the model is not longer applicable in that region. 
To understand the pattern of the characteristic curves we note that,
in the region of interest, the characteristic curves have a universal
shape, i.e., for small values of $\mu$ the shape is independent of
the actual start value $G^2 (T, \mu=0)$.
For sufficiently small values of $G^2(T, \mu=0)$, the characteristic
curves form a nested set of non-intersecting ellipse like curves.
Increasing $G^2(T, \mu=0)$ causes a levelling-off at larger values of  
$\mu$. The levelling-off sets in at smaller values of $\mu$ for
larger values of $G^2 (T, \mu=0)$. This causes the pattern seen in Fig.~3. 
    
Having $G^2(T,\mu)$ at our disposal, the EoS follows from eqs.~(1 - 4)
in a straightforward way.
Fig.~4 shows as example a comparison of our model with the lattice QCD data
from \cite{Allton2} for $N_f = 2$.\footnote{
Here the parameters for the effective coupling in eq.~(5) 
are given by $\lambda = 21.3$,
$T_s = 0.93 T_c$, $a = 427.4$, and $b = 404.2$.}
Again, a fairly well reproduction of the data is achieved
for $T > T_c$, while at $T < T_c$, in particular for large values of $\mu$,
the agreement is less perfect. 

\vskip 6mm
\begin{figure}[ht]
\centering
\includegraphics[width=6.1cm,angle=-90]{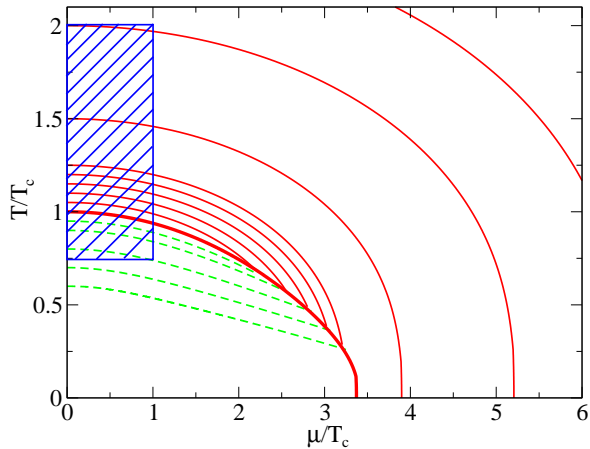}
\caption{
\label{fig:flow}
{\it The characteristic curves solving the flow equation for the effective
coupling $G^2$. Solid (dashed) lines: characteristic curves
emerging from $T > T_c$ ($T< T_c$). The heavy solid curve is considered
as indicator of the phase border line.
The characteristic curves are not drawn in the region
where they intersect.
The marked region indicates where the lattice
QCD data \protect\cite{Allton2} are given. Note that they are actually provided
by the truncated expansion
$\Delta p / T^4 \equiv p(T, \mu=0) / T^4 - p(T, \mu) /T^4
= c_2 (T) (\mu / T)^2 + c_4(T) (\mu / T)^4$. Higher order terms,
like $c_6 (T) \, (\mu / T)^6$ etc.\ are needed to control the validity of
the extrapolation to larger values of $\mu$.}}
\end{figure}
\begin{figure}[ht]
\vskip 3mm
\centering
\includegraphics[width=5.1cm,angle=-90]{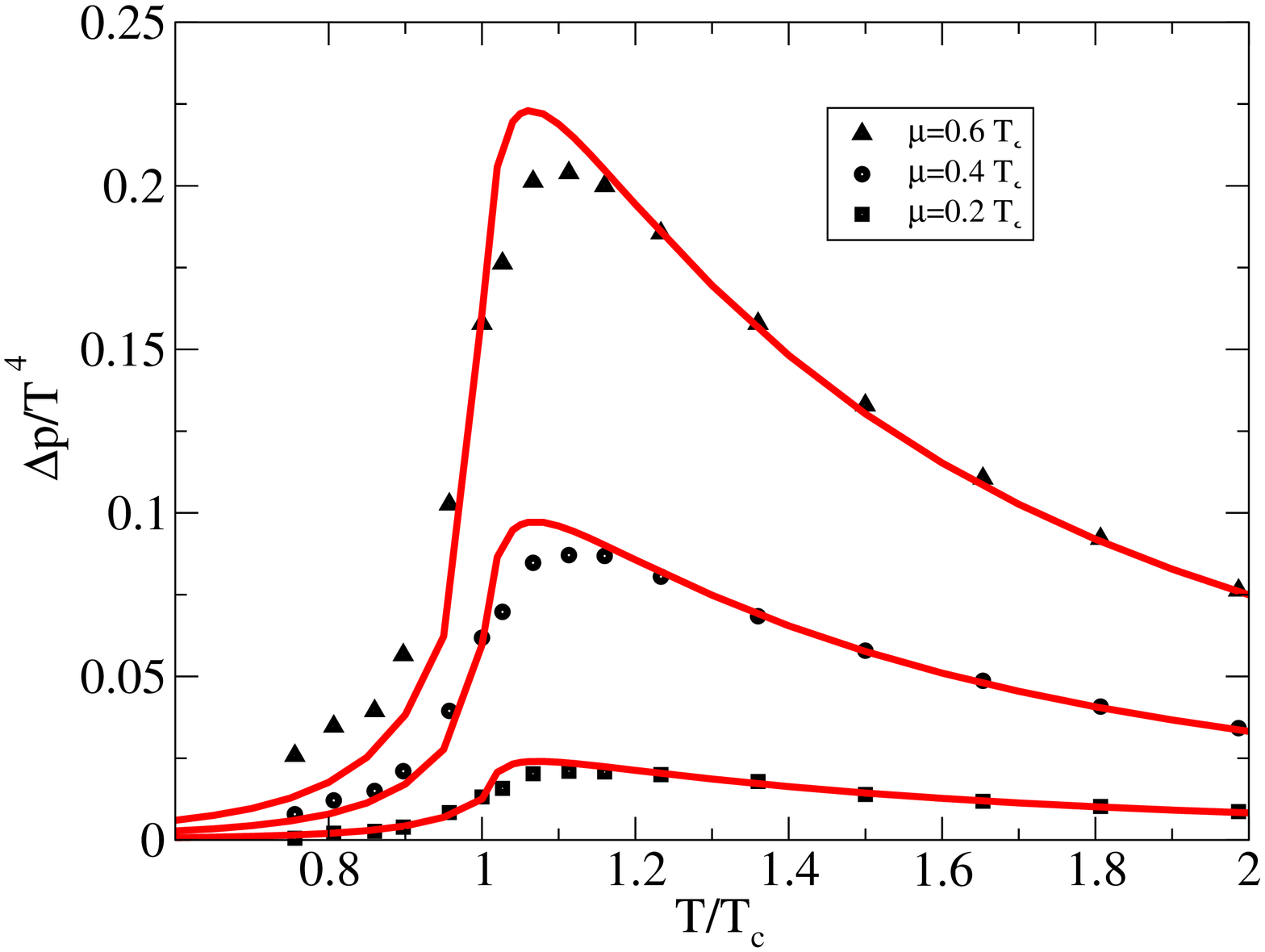}
\hfill
\includegraphics[width=5.1cm,angle=-90]{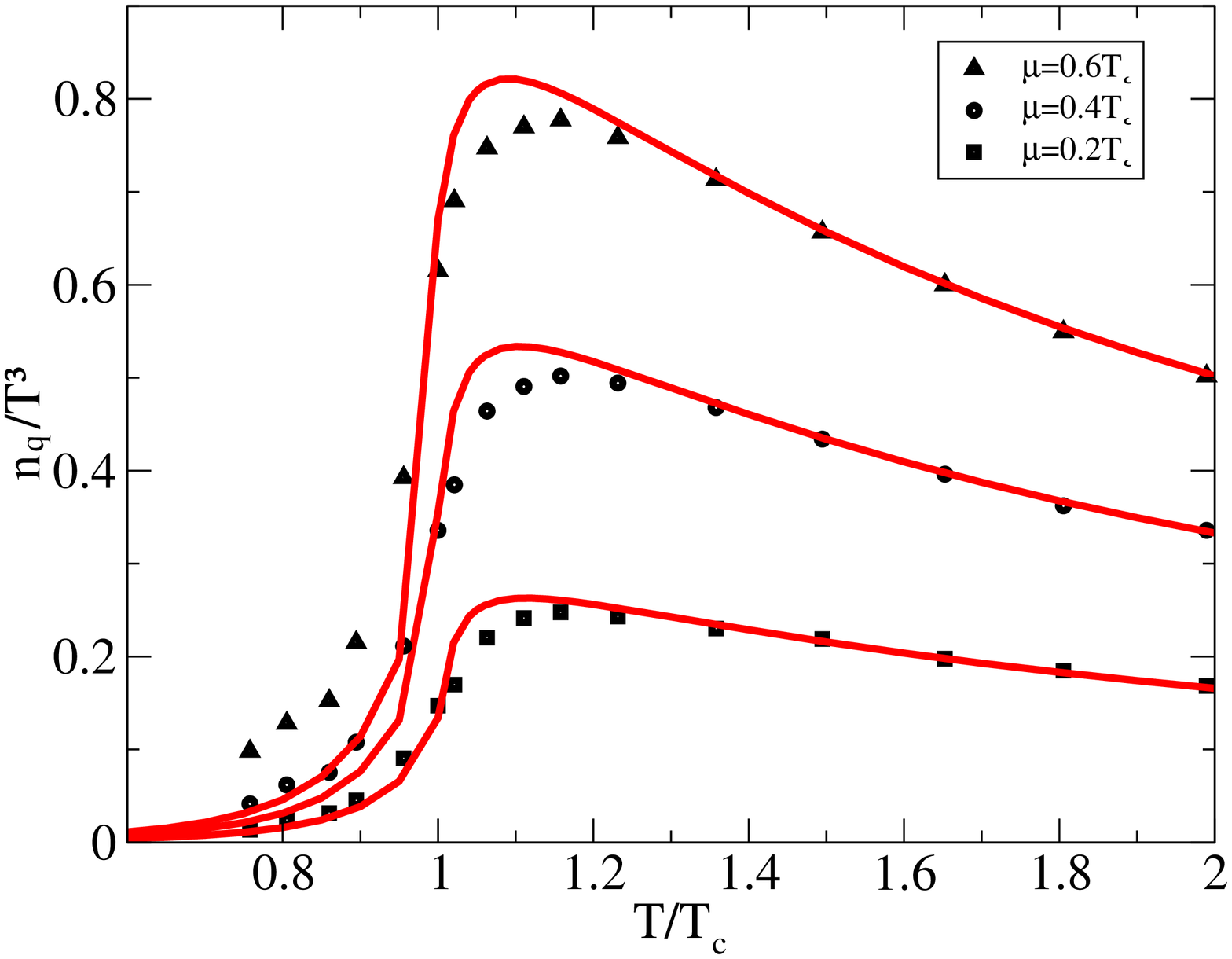}

\caption{
\label{fig:n_q}
{\it The scaled pressure difference $\Delta p (T,\mu) = p(T, \mu) - p(T,\mu = 0)$
(left panel) and the scaled quark number density (right panel) 
as a function of the temperature for various
chemical potentials. Data from \protect\cite{Allton2}.}}
\end{figure}
  
\section{Conclusions} 

To summarize, we present a unique parameterization of the equation of
state of strongly interacting matter in the temperature interval
$0.6 T_c \cdots 3 T_c$  
which is based on the picture of weakly interacting
quasi-particles of quarks and gluons. 
At $T_c$ for $\mu = 0$, the temperature dependence of the effective
coupling suffers a change; no rapidly changing order parameter is
needed to describe the lattice QCD data. Further precision lattice data below
$T_c$ are needed to check the model in more detail and to put the
speculation on firm ground, such that the applicability of the
quasi-particle picture provides an additional example of quark-hadron
duality. 

It has been shown that our model is also able to describe the new lattice QCD data
for non-vanishing chemical potential, as alternative approaches
do \cite{Thaler,Rafelski}. As particular new point we note that
we cover the region $T < T_c$ at
non-vanishing but small chemical potential with our quasi-particle model.

While present lattice QCD data deliver either $p(T, \mu = 0)$ or
$\Delta p(T, \mu)$, our model covers both quantities on the same footing.

Inspiring discussions with J.P. Blaizot, D. Blaschke, A. Peshier, and K. Redlich
are gratefully acknowledged. 
The work is supported by BMBF 06DR121, GSI and FP6-I3 network.

\end{document}